\documentclass[prl,twocolumn,amsmath,amssymb,superscriptaddress]{revtex4-2}
\usepackage{graphicx}
\usepackage{dcolumn}
\usepackage{bm}
\usepackage{textcomp}
\usepackage[colorlinks= true,urlcolor=blue,linkcolor= blue,citecolor=blue,bookmarks=false,pdfstartview=]{hyperref}

\begin{document}
\title{Magnetic order and ballistic spin transport in a sine-Gordon spin chain}
\author{B.\,M. Huddart}
\affiliation{Centre for Materials Physics, Durham University, Durham DH1 3LE, United Kingdom}
\author{M. Gomil\v{s}ek}
\affiliation{Jo\v{z}ef Stefan Institute, Jamova c. 39, SI-1000 Ljubljana, Slovenia}
\affiliation{Centre for Materials Physics, Durham University, Durham DH1 3LE, United Kingdom}
\author{T.\,J. Hicken}
\affiliation{Centre for Materials Physics, Durham University, Durham DH1 3LE, United Kingdom}
\author{F.\,L. Pratt}
\affiliation{ISIS Facility, STFC Rutherford Appleton Laboratory, Didcot OX11 0QX, United Kingdom}
\author{S.\,J. Blundell}
\affiliation{Department of Physics, Clarendon Laboratory, Oxford University, Parks Road, Oxford OX1 3PU, United Kingdom}
\author{P.\,A. Goddard}
\affiliation{Department of Physics, University of Warwick, Coventry CV4 7AL, United Kingdom}
\author{S. J. Kaech}
\affiliation{Department of Chemistry and Biochemistry, Eastern Washington University, Cheney, Washington 99004, USA}
\author{J.\,L. Manson}
\affiliation{Department of Chemistry and Biochemistry, Eastern Washington University, Cheney, Washington 99004, USA}
\author{T. Lancaster}
\affiliation{Centre for Materials Physics, Durham University, Durham DH1 3LE, United Kingdom}
\date{\today}
\begin{abstract}
We report the results of muon-spin spectroscopy ($\mu^+$SR) measurements on the staggered molecular spin chain [pym-Cu(NO$_3$)$_2$(H$_2$O)$_2$] (pym = pyrimidine), a material previously described using sine-Gordon field theory.  Zero-field $\mu^+$SR reveals a long range magnetically-ordered ground state below a transition temperature $T_\mathrm{N}=0.22(1)$~K.  Using longitudinal-field (LF) $\mu^+$SR we investigate the dynamic response in applied magnetic fields $0< B < 500$~mT and find evidence for ballistic spin transport. Our LF $\mu^+$SR measurements on the chiral spin chain [Cu(pym)(H$_2$O)$_4$]SiF$_6 \cdot $H$_2$O instead demonstrate one-dimensional spin diffusion and the distinct spin transport in these two systems likely reflects differences in their magnetic excitations.  
\end{abstract}

\maketitle

%\section{Introduction}
At low temperatures, the one-dimensional (1D) antiferromagnetic (AF) spin chain hosts a range of exotic
magnetic phenomena including quantum-critical fluctuations, emergent energy gaps and topological excitations. The ideal $S = 1/2$ AF Heisenberg chain has a gapless excitation spectrum, but is highly
sensitive to small modifications. A particularly dramatic effect results from the spins enjoying an
alternating local environment, which is achieved in a so-called staggered spin chain \cite{giamarchi}. This system
hosts a magnetic field-induced gapped phase described by sine-Gordon (SG) quantum-field theory, which predicts
a complex excitation spectrum including solitons, antisolitons and soliton-antisoliton bound states, known as {\it breathers}. Generalized hydrodynamic approaches show that transport in the SG field theory is ballistic, except in certain limits (including low temperature), where the semiclassical result predicting either ballistic or diffusive transport is recovered \cite{bertini}.  However, owing to the scarcity of model material systems, the nature of the spin transport has not been firmly established in experimental realizations of this model. In this Letter, we investigate the staggered spin chain material [pym-Cu(NO$_3$)$_2$(H$_2$O)$_2$] (pym~=~pyrimidine~=~C$_4$H$_4$N$_2$), hereafter Cu-PM, using muon-spin spectroscopy ($\mu^+$SR). Cu-PM is one of relatively few experimental realizations of a staggered spin chain that is well-described by SG theory \cite{dender,feyerherm} and here we reveal the material's ground state and the character of its spin transport. 

In Cu-PM, the Cu$^{2+}$ ions form chains, with the primary magnetic exchange [$J=36.3(5)$~K \cite{wolter}] being mediated by linking pym ligands \cite{yasui}.
% The Cu$^{2+}$ ions occupy a distorted octahedral environment, with the principal axes of these octahedra being tilted from the $ac$ plane by %$29.4^{\circ}$.  This axis almost coincides with a principal axis of the $g$ tensor and results in an alternating $g$ tensor for Cu$^{2+}$ ions %along the chain \cite{feyerherm}.
It has been shown theoretically \cite{oshikawa1} that the material's staggered local $g$ tensor of neighboring Cu$^{2+}$ ions produces an effective staggered internal magnetic field transverse but proportional in magnitude to the applied magnetic field. This internal field, which can also be produced by alternating Dzyaloshinskii-Moriya (DM) interactions along the chain, results in a magnetic field-induced gap $\Delta$, which has been experimentally observed for Cu-PM \cite{feyerherm,wolter}. In addition, signatures of the three lowest breathers and a soliton predicted by SG theory have been observed with electron-spin resonance (ESR) \cite{zvyagin}. In the perturbative spinon regime, $\Delta < T < J$,  the soliton-breather superstructure is suppressed and the anisotropic contribution to the Hamiltonian due to the staggered $g$ tensors and DM interactions can be treated a perturbation. This perturbation is predicted to result in a field shift and broadening of the ESR lineshape \cite{oshikawa2}, and the measured ESR parameters for Cu-PM show excellent quantitative agreement with these predictions \cite{zvyagin2}. 

While the subject of zero-field (ZF) magnetic order has been investigated in other SG spin chains, the magnetic ground state of Cu-PM has not yet been established. Despite earlier claims that anomalies in the temperature dependence of the ESR frequency \cite{oshima} and magnetic susceptibility \cite{dender} of Cu-benzoate [Cu(C$_6$D$_5$OO)$_2$ $\cdot$ 3D$_2$O] at $T \approx 0.8$~K were due to an AF phase transition, no evidence for N\'{e}el ordering was found from neutron scattering experiments \cite{dender2}, and it was later argued that the ESR peak instead corresponds to an SG  breather excitation \cite{oshikawa2}. A subsequent $\mu^+$SR study found no evidence of long-range order (LRO) down to 20 mK \cite{asano}. On the other hand, CuCl$_2\cdot$ 2DMSO (CDC) has been shown to acquire zero-field LRO below $T_\mathrm{N}=0.93$~K \cite{chen}. The interchain interactions responsible for this LRO lead to significant deviations from the SG model predictions below the $B_\mathrm{c}=3.9$~T critical field for N\'{e}el order \cite{kenzelmann}.
%Muons are extremely sensitive probes for the detection of long-range magnetic order (LRO), as shown in our previous measurements on molecular magnets \cite{manson,goddard,manson2,dossantos,xiao,lancaster-prl}. 
In this study, we use $\mu^+$SR to show that the ground state of Cu-PM is long-range ordered at temperatures below $T_\mathrm{N}=0.22(1)$~K.
We determine muon stopping sites using density functional theory (DFT) and use these to provide further insight into the nature of the ordered moments. We also use  $\mu^{+}$SR to explore spin transport above $T_\mathrm{N}$ in the perturbative spinon regime and show that it is ballistic at all  measured temperatures, and contrast these with measurements on chiral staggered spin chain [Cu(pym)(H$_2$O)$_4$]SiF$_6$$\cdot$H$_2$O, which instead demonstrate diffusive transport. 

\begin{figure}[t]
	\centering
	\includegraphics[width=\columnwidth]{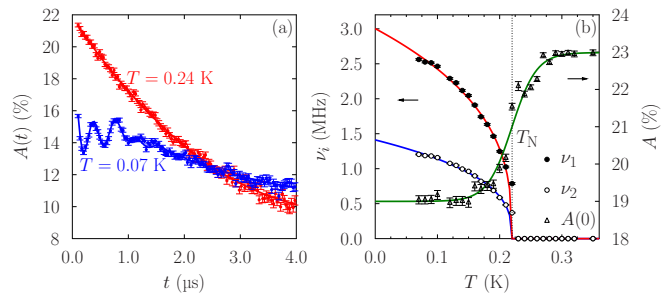}
	\caption{(a) Example ZF spectra above and below the magnetic ordering temperature. (b) Temperature dependence of the (\textit{left axis}) precession frequencies and (\textit{right axis}) initial asymmetry.}
	\label{fig:zf}
\end{figure}

%\section{Experimental}
\footnotetext{See Supplemental Material at [URL will be inserted by publisher] for details of the $\mu^+$SR measurements and the density functional theory and dipolar field calculations.}
%\section{Results}
ZF $\mu^+$SR measurements \cite{steve_review,Note1} reveal a state of LRO in Cu-PM below $T_\mathrm{N}=0.22(1)$~K, indicated by the appearance of spontaneous oscillations at two frequencies [see Fig.~\ref{fig:zf}(a)].  The spectra for $ T \le 0.22$~K were therefore fitted to
%\begin{multline}\label{eq1}
$ A(t)=\sum_{i=1}^2 A_ie^{-\lambda_i t}\cos(2\pi\nu_i t )+A_3 e^{-\sigma^2 t^2/2}
 +A_\mathrm{bg} e^{-\lambda_\mathrm{bg}t},$
%\end{multline}
where the components with amplitudes $A_1$ and $A_2$ account for muons stopping in the sample at two magnetically distinct sites and undergoing coherent spin precession with frequencies $\nu_i$ and relaxation rates $\lambda_i$. An additional component $A_3$ with Gaussian relaxation arises from muons occupying a third distinct site that are sensitive mainly to quasistatic nuclear fields. Muons stopping outside of the sample or those with their spins aligned parallel to the local field contribute to a slowly-relaxing background with amplitude $A_\mathrm{bg}$ and relaxation rate $\lambda_\mathrm{bg}$. The amplitudes $A_1$, $A_2$ and $A_3$ are proportional to the fraction of muons in each distinct magnetic environment and these account for approximately 35\%, 20\% and 45\% of the relaxing part of the asymmetry, respectively. The frequencies $\nu_1$  and $\nu_2$ were found to vary in fixed proportion, so we fixed $\nu_2=0.47\nu_1$ in the fitting procedure.  The frequencies $\nu_i(T)$, shown in Fig.~\ref{fig:zf}(b), are proportional to the magnetic field at the muon site and act as an order parameter for the system. A fit to the critical scaling function $\nu_1(T)=\nu_1(0)(1-T/T_\mathrm{N})^\beta$, appropriate near a second-order phase transition, yields an estimate $T_\mathrm{N} = 0.22(1)$~K for the ordering temperature and $\beta=0.35(1)$ for the critical exponent. The critical exponent $\beta$ obtained here is close to the value 0.367 of the 3D Heisenberg universality class and not far from 0.327 of the 3D Ising universality class \cite{steve_book}, suggesting that that the ground-state LRO in Cu-PM is three-dimensional in nature. 
%Its value is also significantly higher than the critical exponents found for most other $S=1/2$ AF Heisenberg spin chains, for which a precipitous drop in $\nu$ on approaching $T_\mathrm{N}$ is accompanied by a much smaller value of $\beta$, which reflects the low-dimensional nature of the critical fluctuations in those systems close to their phase transitions \cite{cupyz,cupyz3}. 
The precession frequencies $\nu_1(0)=3.0(1)$~MHz and $\nu_2(0)=1.41(5)$~MHz correspond to local magnetic fields of magnitudes $B_1(0)=22.1(7)$~mT and $B_2(0)=10.4(3)$~mT, respectively, for each of the magnetically distinct muon stopping sites. Above $T_\mathrm{N}$, the spectra can be described by the sum of an exponential and a Gaussian relaxation, with a slowly relaxing background as before; the exponential term is due to fluctuating disordered electronic moments.  The magnetic ordering transition is also detected through a drop in the initial asymmetry as an increasing fraction of the muon-spin ensemble experiences large internal magnetic fields and are rapidly dephased from the spectra.  The temperature-dependence of the initial asymmetry is shown in Fig.~\ref{fig:zf}(b) and exhibits a narrow transition between two approximately constant values, with a fit to a Fermi-like step function yielding the same transition temperature as estimated above.

\begin{figure*}[t]
	\centering
	\includegraphics[width=\textwidth]{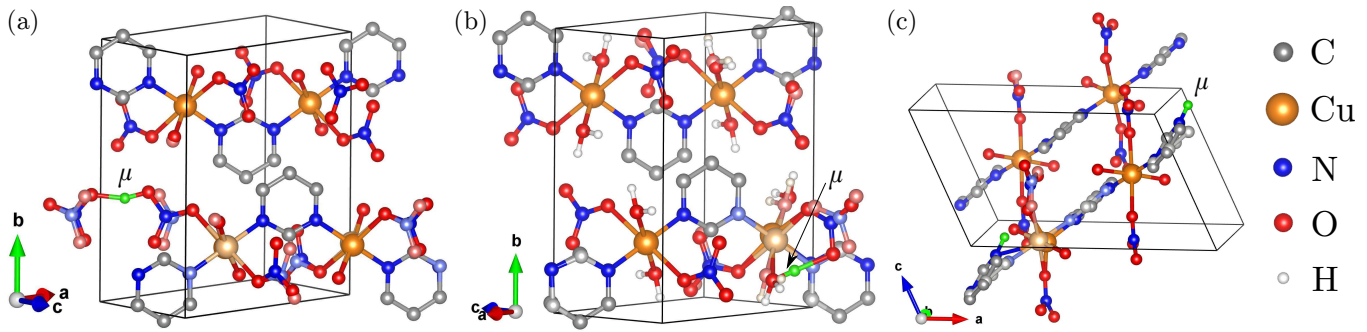}
	\caption{Low-energy muon sites in [pym-Cu(NO$_3$)$_2$(H$_2$O)$_2$]. Lighter spheres represent the ionic positions in the unit cell without
		the muon. H atoms have been omitted for clarity where appropriate. (a) The nitrate site. (b) The H$_2$O site. (c) the N(pym) site.}
	\label{fig:sites}
\end{figure*}

To quantify the ordered moments and any structural distortions due to the presence of a muon, we have carried out density functional theory (DFT) calculations of the muon stopping sites \cite{Note1} using the plane-wave basis-set electronic structure code \textsc{castep} \cite{CASTEP}. We identify three distinct classes of muon stopping site and show these in Fig.~\ref{fig:sites}. Sites where muon sits around 1~\AA\ from an O atom in a nitrate group [Fig.~\ref{fig:sites}(a)] or H$_2$O ligand [Fig.~\ref{fig:sites}(b)] are the lowest and second lowest-energy class of sites, respectively. We also find candidate sites where the muon sits 1.0~\AA~from an N atom in a pym ligand [Fig.~\ref{fig:sites}(c)], which are substantially higher in energy and result in larger local distortions to the crystal structure.  These sites can be mapped to features in the ZF spectra by considering the dipolar fields resulting from candidate AF magnetic structures. From our dipolar field calculations \cite{Note1} we obtain fields of 9--40~mT/$\mu_\mathrm{Cu}$ for the nitrate site, 57--63~mT/$\mu_\mathrm{Cu}$ for the H$_2$O site and 93--99~mT/$\mu_\mathrm{Cu}$ for the N(pym) site, where $\mu_\mathrm{Cu}$ is the ordered moment of the Cu$^{2+}$ ions in Bohr magnetons. We note that the relative size of the calculated fields for the H$_2$O and nitrate sites is consistent with the ratio between frequencies $\nu_1$ and $\nu_2$, respectively. This assignment gives an estimate $\mu_\mathrm{Cu} \approx 0.37\mu_\mathrm{B}$ for the ordered moment. This moment size would give a precession frequency of around 5 MHz for the N(pym) site, which should be resolvable. However, given the fact that this site is  0.4 eV higher in energy than the other two, it is probably not realized in practice, suggesting that the actual muon sites are associated with only small local distortions of the structure. 

Although the ideal 1D $S=1/2$ quantum Heisenberg AF is not expected to show long-range ordering for $T > 0$ \cite{sachdev}, experimental realizations of this model are found to order at low, but non-zero temperature due to the presence of interactions between chains.
A useful figure of merit for the degree to which low-dimensionality is achieved is the ratio $T_\mathrm{N}/J$, as this quantity should be zero for the ideal case and close to unity for an isotropic material.  This quantity can also be used to estimate the interchain coupling $J'$ using a formula obtained from the results of Quantum Monte Carlo calculations on AF chains \cite{QMC}. For Cu-PM we obtain $T_\mathrm{N}/J=6.1 \times 10^{-3}$, $J' \approx 0.09$~K and $|J'/J|=2.5 \times 10^{-3}$, though we note that additional terms present in the Hamiltonian for a staggered spin chain are likely to lower $T_\mathrm{N}$ below the value expected for a simple 1D AF chain and therefore this estimate of $J'$ serves as a lower bound.
% Values of $T_\mathrm{N}$, $J$, $T_\mathrm{N}/J$ and  $|J'/J|$ are shown for Cu-PM and a selection of other Cu$^{2+}$ spin chains in
We can also use these parameters to estimate the size of the ordered moment, using the formula $m \approx 2.034| J'/J|^{1/2}$ obtained from a model of weakly coupled AF spin chains \cite{schulz}. For Cu-PM this yields $m \approx 0.1\mu_\mathrm{B}$, demonstrating that the ordered moment is heavily renormalized due to enhanced quantum fluctuations in this low-dimensional system, an effect that is also seen, to a lesser extent, in our estimate $\mu_\mathrm{Cu} \approx 0.37\mu_\mathrm{B}$ obtained by considering the dipolar field at the muon sites. 

\begin{table}[t]
	\centering
	\caption{\label{table1}Ordering temperature $T_\mathrm{N}$ and intrachain exchange $J$ for a series of Cu-chain compounds. Note that Cu(pyz)(NO$_3$)$_2$ is a linear chain, whereas the rest are staggered chains.}
	\resizebox{\columnwidth}{!}{
		\begin{tabular}{lccc}
			\hline
			\hline
			& $T_\mathrm{N}$ (K) & $J$ (K) & $T_\mathrm{N}/J$\\
			\hline
			Cu(pyz)(NO$_3$)$_2$ \cite{cupyz,hammar} & 0.105(2) & 10.3(1) & $1.0 \times 10^{-2}$\\
			Cu-benzoate \cite{asano,dender} & $< 0.02$ & 18.2(1) & $<10^{-3}$\\
			CDC \cite{chen} & 0.93 & 16.9(1) & $5.5 \times 10^{-2}$\\
			{[Cu(pym)(H$_2$O)$_4$]SiF$_6$$\cdot$H$_2$O} \cite{liu} & $< 0.02$ & 42(1) & $< 5 \times 10^{-4}$\\
			
			Cu-PM \cite{feyerherm,wolter}  & 0.22(1) & 36.3(5) &  $6.1 \times 10^{-3}$\\
			\hline
			\hline
	\end{tabular}}
	
\end{table}

The values of $T_{\mathrm{N}}/J$  (Table~\ref{table1}) suggest that the degree of isolation of the Cu$^{2+}$ chains in Cu-PM is similar to that found in the well-isolated 1D AF linear spin chain copper pyrazine dinitrate [Cu(pyz)(NO$_3$)$_2$] \cite{cupyz}.  %However, the Cu$^{2+}$ ions in Cu(pyz)(NO$_3$)$_2$ do not experience the staggered local environment that they do in Cu-PM.
%Considering instead the staggered spin chain Cu-benzoate [Cu(C$_6$D$_5$OO)$_2\cdot3$D$_2$O] \cite{dender} allows a more direct comparison to be made. We see that with a significantly lower ordering temperature and an intrachain exchange $J$ that is twice as large, showing the better isolation of Cu-PM.
In CDC, which, like Cu-PM, exhibits alternating $g$ tensors and DM interactions, the ground state in zero-field is a collinear AF below $T_\mathrm{N}=0.93$~K, with a moment 0.44(5)$\mu_\mathrm{B}$ along an Ising-like easy axis \cite{chen}. The higher transition temperature in CDC may be due to the lower DM energy in this system \cite{chen}, which is an order of magnitude smaller relative to $J$ than the values established for Cu-PM \cite{feyerherm} and Cu-benzoate \cite{oshikawa1}. 
A more recent example of a staggered chain is the chiral chain [Cu(pym)(H$_2$O)$_4$]SiF$_6$$\cdot$H$_2$O, in which the environments of adjacent Cu$^{2+}$ ions are related by $4_1$ screw symmetry \cite{cordes}. At zero field, its magnetism is well described as a 1D $S = 1/2$ Heisenberg AF, with intrachain exchange $J = 42(1)$~K and no magnetic order detected down to 20~mK \cite{liu}. The suppression of LRO in this system may be due to a uniform DM interaction that alternates in sign from chain to chain, which has been shown theoretically to effectively result in a cancellation of the interchain interaction \cite{starykh}.  
%While it has been the subject of many theoretical studies \cite{starykh,sato,xi,mahdavifar}, the nature of the magnetic ground state in quasi-1D Heisenberg antiferromagnets with staggered $g$ tensors and DM interactions has not yet been firmly established experimentally. Further measurements of the ordered state in Cu-PM using techniques such as neutron diffraction could provide significant insight into this matter.

Longitudinal-field (LF) $\mu^+$SR is often used to investigate low-energy dynamics in 1D materials, including spin chains \cite{xiao} and ladders \cite{spinladder}. The principle of these measurements is to work at a temperature $T$ in the 1D regime, $T_\mathrm{N} \ll T \lesssim J$ (i.e.\ above any ordering temperature, and below the energy scale of the spin exchange along the chain), where collective low-dimensional behavior of the spins is expected. In cases where the isotropic hyperfine coupling $A$ dominates over the dipolar coupling, the field-dependence of the muon-spin relaxation rate $\lambda$ is given by $\lambda(B)=(A^2/4)f(\omega_\mathrm{e})$, where $f(\omega)$ is the spectral density and the probe frequency $\omega_\mathrm{e}=\gamma_\mathrm{e}B$, where $\gamma_\mathrm{e}$ is the electron gyromagnetic ratio \cite{devreux}.
Field-dependent measurements \cite{pratt,xiao} can be used to distinguish between the two main types of spin transport possible in a 1D chain, namely spin diffusion or ballistic transport, since their spin autocorrelation functions have different associated spectral densities: $f(\omega) \propto \omega^{-1/2}$ for diffusive transport and $f(\omega) \propto \ln(cJ/\omega)$ for ballistic motion, where $c$ is a constant of order unity. We have carried out LF $\mu^+$SR measurements on Cu-PM to study the spin dynamics in this system. The data were fitted to
\begin{equation}\label{eq1}
A(t)=A_\mathrm{rel} G_z^\mathrm{KT} (\Lambda,B) e^{-\lambda t}+A_\mathrm{bg}, 
\end{equation}					      
which includes the contribution of the quasistatic nuclear moments through the LF Kubo-Toyabe function $G_z^\mathrm{KT}(\Lambda,B)$, where $\Lambda$ is the width of the resulting field distribution, and the relaxation due to dynamics with relaxation rate $\lambda$.  

\begin{figure}[t]
	\centering
	\includegraphics[width=\columnwidth]{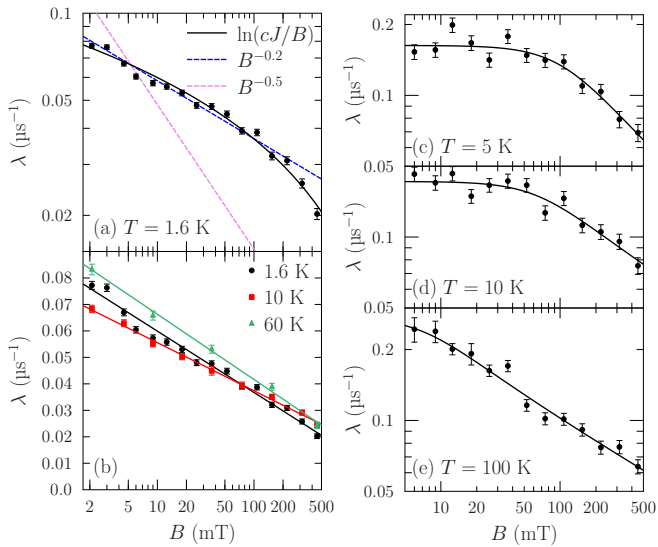}
	\caption{(a) Field dependence of the LF relaxation rate in Cu-PM at $T=1.6$~K with fits appropriate for either diffusive or ballistic spin transport. (b) Field dependence of relaxation rate and fits describing ballistic spin transport at three different temperatures, plotted on a linear--log scale. Field dependence of the LF relaxation rate for [Cu(pym)(H$_2$O)$_4$]SiF$_6 \cdot $H$_2$O at (c) $T= 5$~K, (d) $T=10$~K and (e) $T = 100$~K.}
	\label{fig:lf}
\end{figure}

The field-dependence of the relaxation rate $\lambda$ is shown in Fig.~\ref{fig:lf}. 
%(In the measured field and temperature regime, we expect temperature-driven excitations across the spin gap $\Delta_{\mathrm{s}}$.) 
In Fig.~\ref{fig:lf}(a) we show $\lambda(B)$ at $T=1.6$~K, alongside fits obtained from three different models.  A fit to a function of the form $\lambda \propto B^{-n}$, appropriate for diffusive spin transport, yields $n=0.2$, which is much smaller than the theoretically predicted value of $n=0.5$ for one-dimensional diffusion.  As seen in Fig.~\ref{fig:lf}(a), a function of this form with $n=0.5$ does not describe the data.  However, a function of the form $\lambda(B) = a\ln(cJ/B)$, appropriate for ballistic transport, provides a good fit to the data with $c=0.16$ and outperforms the (unphysical) $\lambda \propto B^{-0.2}$ power-law model at higher fields, where $\lambda(B)$ obtained from a power-law does not drop off quickly enough with increasing field. 
In Fig.~\ref{fig:lf}(b), we show the results of fitting the ballistic model to the data measured at several different temperatures; the model provides a good fit to the data at all of the measured temperatures and even up to $T \approx 2J$.
%(with $c=0.48$ and $c=0.20$ for the fits to the data at 10~K and 100~K, respectively). 

We also carried out LF $\mu^+$SR measurements on the chiral staggered spin chain [Cu(pym)(H$_2$O)$_4$]SiF$_6 \cdot $H$_2$O \cite{liu} to compare the spin dynamics with those measured for Cu-PM. The measured spectra were fitted to the function in Eq.~\eqref{eq1} and we show the field-dependent relaxation rates $\lambda$ in Figs.~\ref{fig:lf}(c-e). As seen in Fig. \ref{fig:lf}(c), the relaxation rate $\lambda(B)$ is approximately constant for fields up to around 50~mT at $T=5$~K. We account for this behavior using a model for anisotropic spin diffusion, where the spectral density has the form
\begin{equation} \label{eq2}
f(\omega)=\frac{1}{\sqrt{2D_\parallel D_\perp }}\left(\frac{1+\sqrt{1+(\omega/2D_\perp)^2}}{2[1+(\omega/2D_\perp)^2]}\right)^n,
\end{equation}
with $D_\parallel$ and $D_\perp$ being the intrachain and interchain diffusion rates, respectively \cite{mizoguchi}. This function shows a transition from a constant low frequency value to a power-law behavior $f(\omega) \propto \omega^{-n}$ at a crossover frequency proportional to $D_\perp$. (This model does not work well for Cu-PM, as there $\lambda(B)$ is not approximately constant at low fields.) Fitting the data measured at 5 K to this model we obtain $D_\perp=10^{10}$~s$^{-1}$ and $n=0.46(12)$. This value of $n$ is in good agreement with the theoretical prediction $n=0.5$. As the temperature is raised we observe a shift in the crossover to lower fields [see Figs.~\ref{fig:lf}(d,e)], corresponding to a decrease in $D_\perp$ and we also find that $D_\parallel$ increases with increasing temperature. 
%(though obtaining absolute values of $D_\parallel$ would require knowledge of the hyperfine coupling $A$). 
We note that the fits to the data at 10~K [Fig.~\ref{fig:lf}(d)] and 100~K [Fig.~\ref{fig:lf}(e)]  yield values of $n$ closer to 0.3, which, although noticeably smaller than theoretically predicted, are not too dissimilar from the value $n \approx 0.35$ measured for the linear chain DEOCC-TCNQF$_4$ \cite{pratt}. Furthermore, we would expect the model in Eq. \eqref{eq2} to be most accurate at $T=5$~K because, as the temperature approaches (or exceeds) $ T \approx J$, the muons are responding not only to delocalized spin excitations but also to the quasi-independent spin flips introduced by thermal fluctuations.

Both ballistic \cite{e184404,maeter} and diffusive \cite{thurber,pratt,xiao} spin transport have previously been observed experimentally in $S=1/2$ AF spin chains, although the latter is far more common. The nature of spin transport in these systems remains controversial, with recent theoretical work suggesting that, in the presence of a periodic lattice potential, diffusion can coexist with ballistic transport \cite{sirker}. For the vast majority of the data measured for Cu-PM we are within the perturbative spinon regime, where the soliton and breather modes are suppressed \cite{zvyagin2}. It is therefore more likely that the excitations responsible for the observed spin transport are spinons rather than solitons.   The anisotropic term in the Hamiltonian due to the staggered $g$ tensor and DM interaction is likely to modify the spectral density of the spin excitations compared with those found in other $S=1/2$ AF spin chains and could therefore be responsible for their distinct transport. There is evidence for other excitations, such as interbreather transitions \cite{tiegel}, in this intermediate temperature regime, which could also influence the spin transport. The transport in [Cu(pym)(H$_2$O)$_4$]SiF$_6 \cdot $H$_2$O is very different, despite both systems being expected to occupy a similar regime of behavior (i.e where the temperature exceeds the spin gap, but is smaller than the exchange strength $J$). However, despite the presence of staggered $g$ tensors, a staggered susceptibility and a spin gap that opens on the application of a magnetic field \cite{liu}, all of which are reminiscent of phenomena observed in non-chiral staggered chains, the size of the gap in [Cu(pym)(H$_2$O)$_4$]SiF$_6 \cdot $H$_2$O and its linear field dependence do not fit with the predictions of the SG model. These differences were attributed to additional anistropic interactions \cite{liu}, and it is possible that these interactions modify the character of the excitations in this regime, which in turn alters the nature of their transport. 

In conclusion, we have demonstrated the existence of long-range magnetic order in the staggered spin chain compound [pym-Cu(NO$_3$)$_2$(H$_2$O)$_2$] with $T_\mathrm{N}=0.22(1)$~K.  Our LF $\mu^+$SR measurements show that the transport of the spin excitations detected by the muon is ballistic in the perturbative spinon regime, $\Delta < T < J$, whereas the transport in the chiral spin chain [Cu(pym)(H$_2$O)$_4$]SiF$_6 \cdot $H$_2$O is shown to be diffusive. This difference is taken to reflect the distinct character of the excitations in each of these systems. Establishing the sensitivity of the
muon to the magnetic excitations in these systems paves the way for $\mu^+$SR measurements at higher magnetic fields, where implanted muons could provide insight into the soliton-breather regime $T \ll \Delta$.  

Experiments were carried out at the STFC-ISIS Facility and we are grateful for the provision of beamtime. We thank J. Singleton and J.\,L. Musfeldt for useful discussion. This work is supported by EPSRC (UK), under grants EP/N024028/1 and EP/N032128/1.
We acknowledge computing resources provided by Durham Hamilton HPC. MG acknowledges support by the Slovenian Research Agency under project No. Z1-1852. Work at EWU was supported by the U.S. National Science Foundation under grant no. DMR-1703003. This project has received funding from the European Research Council (ERC) under the European Union’s Horizon 2020 research and innovation program (Grant Agreement No. 681260). Research data will be made available via [INSERT http LINK].


\begin{thebibliography}{37}%
	\makeatletter
	\providecommand \@ifxundefined [1]{%
		\@ifx{#1\undefined}
	}%
	\providecommand \@ifnum [1]{%
		\ifnum #1\expandafter \@firstoftwo
		\else \expandafter \@secondoftwo
		\fi
	}%
	\providecommand \@ifx [1]{%
		\ifx #1\expandafter \@firstoftwo
		\else \expandafter \@secondoftwo
		\fi
	}%
	\providecommand \natexlab [1]{#1}%
	\providecommand \enquote  [1]{``#1''}%
	\providecommand \bibnamefont  [1]{#1}%
	\providecommand \bibfnamefont [1]{#1}%
	\providecommand \citenamefont [1]{#1}%
	\providecommand \href@noop [0]{\@secondoftwo}%
	\providecommand \href [0]{\begingroup \@sanitize@url \@href}%
	\providecommand \@href[1]{\@@startlink{#1}\@@href}%
	\providecommand \@@href[1]{\endgroup#1\@@endlink}%
	\providecommand \@sanitize@url [0]{\catcode `\\12\catcode `\$12\catcode
		`\&12\catcode `\#12\catcode `\^12\catcode `\_12\catcode `\%12\relax}%
	\providecommand \@@startlink[1]{}%
	\providecommand \@@endlink[0]{}%
	\providecommand \url  [0]{\begingroup\@sanitize@url \@url }%
	\providecommand \@url [1]{\endgroup\@href {#1}{\urlprefix }}%
	\providecommand \urlprefix  [0]{URL }%
	\providecommand \Eprint [0]{\href }%
	\providecommand \doibase [0]{https://doi.org/}%
	\providecommand \selectlanguage [0]{\@gobble}%
	\providecommand \bibinfo  [0]{\@secondoftwo}%
	\providecommand \bibfield  [0]{\@secondoftwo}%
	\providecommand \translation [1]{[#1]}%
	\providecommand \BibitemOpen [0]{}%
	\providecommand \bibitemStop [0]{}%
	\providecommand \bibitemNoStop [0]{.\EOS\space}%
	\providecommand \EOS [0]{\spacefactor3000\relax}%
	\providecommand \BibitemShut  [1]{\csname bibitem#1\endcsname}%
	\let\auto@bib@innerbib\@empty
	%</preamble>
	\bibitem [{\citenamefont {Giamarchi}(2004)}]{giamarchi}%
	\BibitemOpen
	\bibfield  {author} {\bibinfo {author} {\bibfnamefont {T.}~\bibnamefont
			{Giamarchi}},\ }\href
	{https://doi.org/10.1093/acprof:oso/9780198525004.001.0001} {\emph {\bibinfo
			{title} {{Quantum physics in one dimension}}}}\ (\bibinfo  {publisher}
	{Clarendon Press},\ \bibinfo {address} {Oxford},\ \bibinfo {year}
	{2004})\BibitemShut {NoStop}%
	\bibitem [{\citenamefont {Bertini}\ \emph {et~al.}(2019)\citenamefont
		{Bertini}, \citenamefont {Piroli},\ and\ \citenamefont {Kormos}}]{bertini}%
	\BibitemOpen
	\bibfield  {author} {\bibinfo {author} {\bibfnamefont {B.}~\bibnamefont
			{Bertini}}, \bibinfo {author} {\bibfnamefont {L.}~\bibnamefont {Piroli}},\
		and\ \bibinfo {author} {\bibfnamefont {M.}~\bibnamefont {Kormos}},\
	}\bibfield  {title} {\bibinfo {title} {Transport in the sine-{Gordon} field
			theory: From generalized hydrodynamics to semiclassics},\ }\href
	{https://doi.org/10.1103/PhysRevB.100.035108} {\bibfield  {journal} {\bibinfo
			{journal} {Phys. Rev. B}\ }\textbf {\bibinfo {volume} {100}},\ \bibinfo
		{pages} {035108} (\bibinfo {year} {2019})}\BibitemShut {NoStop}%
	\bibitem [{\citenamefont {Dender}\ \emph {et~al.}(1996)\citenamefont {Dender},
		\citenamefont {Davidovi\ifmmode~\acute{c}\else \'{c}\fi{}}, \citenamefont
		{Reich}, \citenamefont {Broholm}, \citenamefont {Lefmann},\ and\
		\citenamefont {Aeppli}}]{dender}%
	\BibitemOpen
	\bibfield  {author} {\bibinfo {author} {\bibfnamefont {D.~C.}\ \bibnamefont
			{Dender}}, \bibinfo {author} {\bibfnamefont {D.}~\bibnamefont
			{Davidovi\ifmmode~\acute{c}\else \'{c}\fi{}}}, \bibinfo {author}
		{\bibfnamefont {D.~H.}\ \bibnamefont {Reich}}, \bibinfo {author}
		{\bibfnamefont {C.}~\bibnamefont {Broholm}}, \bibinfo {author} {\bibfnamefont
			{K.}~\bibnamefont {Lefmann}},\ and\ \bibinfo {author} {\bibfnamefont
			{G.}~\bibnamefont {Aeppli}},\ }\bibfield  {title} {\bibinfo {title} {Magnetic
			properties of a quasi-one-dimensional {$S$}=1/2 antiferromagnet: Copper
			benzoate},\ }\href {https://doi.org/10.1103/PhysRevB.53.2583} {\bibfield
		{journal} {\bibinfo  {journal} {Phys. Rev. B}\ }\textbf {\bibinfo {volume}
			{53}},\ \bibinfo {pages} {2583} (\bibinfo {year} {1996})}\BibitemShut
	{NoStop}%
	\bibitem [{\citenamefont {Feyerherm}\ \emph {et~al.}(2000)\citenamefont
		{Feyerherm}, \citenamefont {Abens}, \citenamefont {Günther}, \citenamefont
		{Ishida}, \citenamefont {Mei{\ss}ner}, \citenamefont {Meschke}, \citenamefont
		{Nogami},\ and\ \citenamefont {Steiner}}]{feyerherm}%
	\BibitemOpen
	\bibfield  {author} {\bibinfo {author} {\bibfnamefont {R.}~\bibnamefont
			{Feyerherm}}, \bibinfo {author} {\bibfnamefont {S.}~\bibnamefont {Abens}},
		\bibinfo {author} {\bibfnamefont {D.}~\bibnamefont {Günther}}, \bibinfo
		{author} {\bibfnamefont {T.}~\bibnamefont {Ishida}}, \bibinfo {author}
		{\bibfnamefont {M.}~\bibnamefont {Mei{\ss}ner}}, \bibinfo {author}
		{\bibfnamefont {M.}~\bibnamefont {Meschke}}, \bibinfo {author} {\bibfnamefont
			{T.}~\bibnamefont {Nogami}},\ and\ \bibinfo {author} {\bibfnamefont
			{M.}~\bibnamefont {Steiner}},\ }\bibfield  {title} {\bibinfo {title}
		{Magnetic-field induced gap and staggered susceptibility in {the \textit{S}}=
			1/2 chain {[{PM}$\cdotp$Cu({NO}$_3$)$_2\cdotp$(H$_2$O)$_2$]$_n$} ({PM} =
			pyrimidine)},\ }\href {https://doi.org/10.1088/0953-8984/12/39/312}
	{\bibfield  {journal} {\bibinfo  {journal} {J. Phys. Condens. Matter}\
		}\textbf {\bibinfo {volume} {12}},\ \bibinfo {pages} {8495} (\bibinfo {year}
		{2000})}\BibitemShut {NoStop}%
	\bibitem [{\citenamefont {Wolter}\ \emph {et~al.}(2003)\citenamefont {Wolter},
		\citenamefont {Rakoto}, \citenamefont {Costes}, \citenamefont {Honecker},
		\citenamefont {Brenig}, \citenamefont {Kl\"umper}, \citenamefont {Klauss},
		\citenamefont {Litterst}, \citenamefont {Feyerherm}, \citenamefont
		{J\'erome},\ and\ \citenamefont {S\"ullow}}]{wolter}%
	\BibitemOpen
	\bibfield  {author} {\bibinfo {author} {\bibfnamefont {A.~U.~B.}\
			\bibnamefont {Wolter}}, \bibinfo {author} {\bibfnamefont {H.}~\bibnamefont
			{Rakoto}}, \bibinfo {author} {\bibfnamefont {M.}~\bibnamefont {Costes}},
		\bibinfo {author} {\bibfnamefont {A.}~\bibnamefont {Honecker}}, \bibinfo
		{author} {\bibfnamefont {W.}~\bibnamefont {Brenig}}, \bibinfo {author}
		{\bibfnamefont {A.}~\bibnamefont {Kl\"umper}}, \bibinfo {author}
		{\bibfnamefont {H.-H.}\ \bibnamefont {Klauss}}, \bibinfo {author}
		{\bibfnamefont {F.~J.}\ \bibnamefont {Litterst}}, \bibinfo {author}
		{\bibfnamefont {R.}~\bibnamefont {Feyerherm}}, \bibinfo {author}
		{\bibfnamefont {D.}~\bibnamefont {J\'erome}},\ and\ \bibinfo {author}
		{\bibfnamefont {S.}~\bibnamefont {S\"ullow}},\ }\bibfield  {title} {\bibinfo
		{title} {High-field magnetization study of the {$S=\frac{1}{2}$}
			antiferromagnetic {Heisenberg} chain
			{$[\mathrm{PM}\mathrm{Cu}({\mathrm{NO}}_{3}{)}_{2}({\mathrm{H}}_{2}\mathrm{O}{)}_{2}{]}_{n}$}
			with a field-induced gap},\ }\href
	{https://doi.org/10.1103/PhysRevB.68.220406} {\bibfield  {journal} {\bibinfo
			{journal} {Phys. Rev. B}\ }\textbf {\bibinfo {volume} {68}},\ \bibinfo
		{pages} {220406(R)} (\bibinfo {year} {2003})}\BibitemShut {NoStop}%
	\bibitem [{\citenamefont {Yasui}\ \emph {et~al.}(2001)\citenamefont {Yasui},
		\citenamefont {Ishikawa}, \citenamefont {Akiyama}, \citenamefont {Ishida},
		\citenamefont {Nogami},\ and\ \citenamefont {Iwasaki}}]{yasui}%
	\BibitemOpen
	\bibfield  {author} {\bibinfo {author} {\bibfnamefont {M.}~\bibnamefont
			{Yasui}}, \bibinfo {author} {\bibfnamefont {Y.}~\bibnamefont {Ishikawa}},
		\bibinfo {author} {\bibfnamefont {N.}~\bibnamefont {Akiyama}}, \bibinfo
		{author} {\bibfnamefont {T.}~\bibnamefont {Ishida}}, \bibinfo {author}
		{\bibfnamefont {T.}~\bibnamefont {Nogami}},\ and\ \bibinfo {author}
		{\bibfnamefont {F.}~\bibnamefont {Iwasaki}},\ }\bibfield  {title} {\bibinfo
		{title} {{Dipyrimidine{--}copper(II) dinitrate complexes showing magnetic
				interactions}},\ }\href {https://doi.org/10.1107/S0108768101002737}
	{\bibfield  {journal} {\bibinfo  {journal} {Acta Crystallogr. B}\ }\textbf
		{\bibinfo {volume} {57}},\ \bibinfo {pages} {288} (\bibinfo {year}
		{2001})}\BibitemShut {NoStop}%
	\bibitem [{\citenamefont {Oshikawa}\ and\ \citenamefont
		{Affleck}(1997)}]{oshikawa1}%
	\BibitemOpen
	\bibfield  {author} {\bibinfo {author} {\bibfnamefont {M.}~\bibnamefont
			{Oshikawa}}\ and\ \bibinfo {author} {\bibfnamefont {I.}~\bibnamefont
			{Affleck}},\ }\bibfield  {title} {\bibinfo {title} {Field-induced gap in
			{$S=1/2$} antiferromagnetic chains},\ }\href
	{https://doi.org/10.1103/PhysRevLett.79.2883} {\bibfield  {journal} {\bibinfo
			{journal} {Phys. Rev. Lett.}\ }\textbf {\bibinfo {volume} {79}},\ \bibinfo
		{pages} {2883} (\bibinfo {year} {1997})}\BibitemShut {NoStop}%
	\bibitem [{\citenamefont {Zvyagin}\ \emph {et~al.}(2004)\citenamefont
		{Zvyagin}, \citenamefont {Kolezhuk}, \citenamefont {Krzystek},\ and\
		\citenamefont {Feyerherm}}]{zvyagin}%
	\BibitemOpen
	\bibfield  {author} {\bibinfo {author} {\bibfnamefont {S.~A.}\ \bibnamefont
			{Zvyagin}}, \bibinfo {author} {\bibfnamefont {A.~K.}\ \bibnamefont
			{Kolezhuk}}, \bibinfo {author} {\bibfnamefont {J.}~\bibnamefont {Krzystek}},\
		and\ \bibinfo {author} {\bibfnamefont {R.}~\bibnamefont {Feyerherm}},\
	}\bibfield  {title} {\bibinfo {title} {Excitation hierarchy of the quantum
			sine-{Gordon} spin chain in a strong magnetic field},\ }\href
	{https://doi.org/10.1103/PhysRevLett.93.027201} {\bibfield  {journal}
		{\bibinfo  {journal} {Phys. Rev. Lett.}\ }\textbf {\bibinfo {volume} {93}},\
		\bibinfo {pages} {027201} (\bibinfo {year} {2004})}\BibitemShut {NoStop}%
	\bibitem [{\citenamefont {Oshikawa}\ and\ \citenamefont
		{Affleck}(1999)}]{oshikawa2}%
	\BibitemOpen
	\bibfield  {author} {\bibinfo {author} {\bibfnamefont {M.}~\bibnamefont
			{Oshikawa}}\ and\ \bibinfo {author} {\bibfnamefont {I.}~\bibnamefont
			{Affleck}},\ }\bibfield  {title} {\bibinfo {title} {Low-temperature electron
			spin resonance theory for half-integer spin antiferromagnetic chains},\
	}\href {https://doi.org/10.1103/PhysRevLett.82.5136} {\bibfield  {journal}
		{\bibinfo  {journal} {Phys. Rev. Lett.}\ }\textbf {\bibinfo {volume} {82}},\
		\bibinfo {pages} {5136} (\bibinfo {year} {1999})}\BibitemShut {NoStop}%
	\bibitem [{\citenamefont {Zvyagin}\ \emph {et~al.}(2005)\citenamefont
		{Zvyagin}, \citenamefont {Kolezhuk}, \citenamefont {Krzystek},\ and\
		\citenamefont {Feyerherm}}]{zvyagin2}%
	\BibitemOpen
	\bibfield  {author} {\bibinfo {author} {\bibfnamefont {S.~A.}\ \bibnamefont
			{Zvyagin}}, \bibinfo {author} {\bibfnamefont {A.~K.}\ \bibnamefont
			{Kolezhuk}}, \bibinfo {author} {\bibfnamefont {J.}~\bibnamefont {Krzystek}},\
		and\ \bibinfo {author} {\bibfnamefont {R.}~\bibnamefont {Feyerherm}},\
	}\bibfield  {title} {\bibinfo {title} {Electron spin resonance in
			sine-{Gordon} spin chains in the perturbative spinon regime},\ }\href
	{https://doi.org/10.1103/PhysRevLett.95.017207} {\bibfield  {journal}
		{\bibinfo  {journal} {Phys. Rev. Lett.}\ }\textbf {\bibinfo {volume} {95}},\
		\bibinfo {pages} {017207} (\bibinfo {year} {2005})}\BibitemShut {NoStop}%
	\bibitem [{\citenamefont {Oshima}\ \emph {et~al.}(1978)\citenamefont {Oshima},
		\citenamefont {Okuda},\ and\ \citenamefont {Date}}]{oshima}%
	\BibitemOpen
	\bibfield  {author} {\bibinfo {author} {\bibfnamefont {K.}~\bibnamefont
			{Oshima}}, \bibinfo {author} {\bibfnamefont {K.}~\bibnamefont {Okuda}},\ and\
		\bibinfo {author} {\bibfnamefont {M.}~\bibnamefont {Date}},\ }\bibfield
	{title} {\bibinfo {title} {Antiferromagnetic resonance in copper benzoate
			below 1 {K}},\ }\href {https://doi.org/10.1143/JPSJ.44.757} {\bibfield
		{journal} {\bibinfo  {journal} {J. Phys. Soc. Jpn}\ }\textbf {\bibinfo
			{volume} {44}},\ \bibinfo {pages} {757} (\bibinfo {year} {1978})}\BibitemShut
	{NoStop}%
	\bibitem [{\citenamefont {Dender}\ \emph {et~al.}(1997)\citenamefont {Dender},
		\citenamefont {Hammar}, \citenamefont {Reich}, \citenamefont {Broholm},\ and\
		\citenamefont {Aeppli}}]{dender2}%
	\BibitemOpen
	\bibfield  {author} {\bibinfo {author} {\bibfnamefont {D.~C.}\ \bibnamefont
			{Dender}}, \bibinfo {author} {\bibfnamefont {P.~R.}\ \bibnamefont {Hammar}},
		\bibinfo {author} {\bibfnamefont {D.~H.}\ \bibnamefont {Reich}}, \bibinfo
		{author} {\bibfnamefont {C.}~\bibnamefont {Broholm}},\ and\ \bibinfo {author}
		{\bibfnamefont {G.}~\bibnamefont {Aeppli}},\ }\bibfield  {title} {\bibinfo
		{title} {Direct observation of field-induced incommensurate fluctuations in a
			one-dimensional {$\mathit{S}=1/2$} antiferromagnet},\ }\href
	{https://doi.org/10.1103/PhysRevLett.79.1750} {\bibfield  {journal} {\bibinfo
			{journal} {Phys. Rev. Lett.}\ }\textbf {\bibinfo {volume} {79}},\ \bibinfo
		{pages} {1750} (\bibinfo {year} {1997})}\BibitemShut {NoStop}%
	\bibitem [{\citenamefont {Asano}\ \emph {et~al.}(2002)\citenamefont {Asano},
		\citenamefont {Nojiri}, \citenamefont {Higemoto}, \citenamefont {Koda},
		\citenamefont {Kadono},\ and\ \citenamefont {Ajiro}}]{asano}%
	\BibitemOpen
	\bibfield  {author} {\bibinfo {author} {\bibfnamefont {T.}~\bibnamefont
			{Asano}}, \bibinfo {author} {\bibfnamefont {H.}~\bibnamefont {Nojiri}},
		\bibinfo {author} {\bibfnamefont {W.}~\bibnamefont {Higemoto}}, \bibinfo
		{author} {\bibfnamefont {A.}~\bibnamefont {Koda}}, \bibinfo {author}
		{\bibfnamefont {R.}~\bibnamefont {Kadono}},\ and\ \bibinfo {author}
		{\bibfnamefont {Y.}~\bibnamefont {Ajiro}},\ }\bibfield  {title} {\bibinfo
		{title} {{$\mu$SR} study of {Cu} benzoate at very low temperature –
			existence or nonexistence of long range order in coupled chains –},\ }\href
	{https://doi.org/10.1143/JPSJ.71.594} {\bibfield  {journal} {\bibinfo
			{journal} {J. Phys. Soc. Jpn}\ }\textbf {\bibinfo {volume} {71}},\ \bibinfo
		{pages} {594} (\bibinfo {year} {2002})}\BibitemShut {NoStop}%
	\bibitem [{\citenamefont {Chen}\ \emph {et~al.}(2007)\citenamefont {Chen},
		\citenamefont {Stone}, \citenamefont {Kenzelmann}, \citenamefont {Batista},
		\citenamefont {Reich},\ and\ \citenamefont {Broholm}}]{chen}%
	\BibitemOpen
	\bibfield  {author} {\bibinfo {author} {\bibfnamefont {Y.}~\bibnamefont
			{Chen}}, \bibinfo {author} {\bibfnamefont {M.~B.}\ \bibnamefont {Stone}},
		\bibinfo {author} {\bibfnamefont {M.}~\bibnamefont {Kenzelmann}}, \bibinfo
		{author} {\bibfnamefont {C.~D.}\ \bibnamefont {Batista}}, \bibinfo {author}
		{\bibfnamefont {D.~H.}\ \bibnamefont {Reich}},\ and\ \bibinfo {author}
		{\bibfnamefont {C.}~\bibnamefont {Broholm}},\ }\bibfield  {title} {\bibinfo
		{title} {Phase diagram and spin hamiltonian of weakly-coupled anisotropic
			{$S=\frac{1}{2}$} chains in
			{$\mathrm{Cu}{\mathrm{Cl}}_{2}\cdot2({(\mathrm{C}{\mathrm{D}}_{3})}_{2}\mathrm{S}\mathrm{O})$}},\
	}\href {https://doi.org/10.1103/PhysRevB.75.214409} {\bibfield  {journal}
		{\bibinfo  {journal} {Phys. Rev. B}\ }\textbf {\bibinfo {volume} {75}},\
		\bibinfo {pages} {214409} (\bibinfo {year} {2007})}\BibitemShut {NoStop}%
	\bibitem [{\citenamefont {Kenzelmann}\ \emph {et~al.}(2004)\citenamefont
		{Kenzelmann}, \citenamefont {Chen}, \citenamefont {Broholm}, \citenamefont
		{Reich},\ and\ \citenamefont {Qiu}}]{kenzelmann}%
	\BibitemOpen
	\bibfield  {author} {\bibinfo {author} {\bibfnamefont {M.}~\bibnamefont
			{Kenzelmann}}, \bibinfo {author} {\bibfnamefont {Y.}~\bibnamefont {Chen}},
		\bibinfo {author} {\bibfnamefont {C.}~\bibnamefont {Broholm}}, \bibinfo
		{author} {\bibfnamefont {D.~H.}\ \bibnamefont {Reich}},\ and\ \bibinfo
		{author} {\bibfnamefont {Y.}~\bibnamefont {Qiu}},\ }\bibfield  {title}
	{\bibinfo {title} {Bound spinons in an antiferromagnetic {$S=1/2$} chain with
			a staggered field},\ }\href {https://doi.org/10.1103/PhysRevLett.93.017204}
	{\bibfield  {journal} {\bibinfo  {journal} {Phys. Rev. Lett.}\ }\textbf
		{\bibinfo {volume} {93}},\ \bibinfo {pages} {017204} (\bibinfo {year}
		{2004})}\BibitemShut {NoStop}%
	\bibitem [{\citenamefont {Blundell}(1999)}]{steve_review}%
	\BibitemOpen
	\bibfield  {author} {\bibinfo {author} {\bibfnamefont {S.~J.}\ \bibnamefont
			{Blundell}},\ }\bibfield  {title} {\bibinfo {title} {Spin-polarized muons in
			condensed matter physics},\ }\href {https://doi.org/10.1080/001075199181521}
	{\bibfield  {journal} {\bibinfo  {journal} {Contemp. Phys.}\ }\textbf
		{\bibinfo {volume} {40}},\ \bibinfo {pages} {175} (\bibinfo {year}
		{1999})}\BibitemShut {NoStop}%
	\bibitem [{Note1()}]{Note1}%
	\BibitemOpen
	\bibinfo {note} {See Supplemental Material at [URL will be inserted by
		publisher] for details of the $\mu ^+$SR measurements and the density
		functional theory and dipolar field calculations.}\BibitemShut {Stop}%
	\bibitem [{\citenamefont {Blundell}(2001)}]{steve_book}%
	\BibitemOpen
	\bibfield  {author} {\bibinfo {author} {\bibfnamefont {S.}~\bibnamefont
			{Blundell}},\ }\href@noop {} {\emph {\bibinfo {title} {Magnetism in condensed
				matter}}}\ (\bibinfo  {publisher} {Oxford University Press},\ \bibinfo
	{address} {Oxford},\ \bibinfo {year} {2001})\BibitemShut {NoStop}%
	\bibitem [{\citenamefont {Clark}\ \emph {et~al.}(2005)\citenamefont {Clark},
		\citenamefont {Segall}, \citenamefont {Pickard}, \citenamefont {Hasnip},
		\citenamefont {Probert}, \citenamefont {Refson},\ and\ \citenamefont
		{Payne}}]{CASTEP}%
	\BibitemOpen
	\bibfield  {author} {\bibinfo {author} {\bibfnamefont {S.~J.}\ \bibnamefont
			{Clark}}, \bibinfo {author} {\bibfnamefont {M.~D.}\ \bibnamefont {Segall}},
		\bibinfo {author} {\bibfnamefont {C.~J.}\ \bibnamefont {Pickard}}, \bibinfo
		{author} {\bibfnamefont {P.~J.}\ \bibnamefont {Hasnip}}, \bibinfo {author}
		{\bibfnamefont {M.~I.~J.}\ \bibnamefont {Probert}}, \bibinfo {author}
		{\bibfnamefont {K.}~\bibnamefont {Refson}},\ and\ \bibinfo {author}
		{\bibfnamefont {M.~C.}\ \bibnamefont {Payne}},\ }\bibfield  {title} {\bibinfo
		{title} {First principles methods using {CASTEP}},\ }\href
	{https://doi.org/10.1524/zkri.220.5.567.65075} {\bibfield  {journal}
		{\bibinfo  {journal} {Z. Kristallogr. Cryst. Mater}\ }\textbf {\bibinfo
			{volume} {220}},\ \bibinfo {pages} {567 } (\bibinfo {year}
		{2005})}\BibitemShut {NoStop}%
	\bibitem [{\citenamefont {Sachdev}(2011)}]{sachdev}%
	\BibitemOpen
	\bibfield  {author} {\bibinfo {author} {\bibfnamefont {S.}~\bibnamefont
			{Sachdev}},\ }\href {https://doi.org/10.1017/CBO9780511973765} {\emph
		{\bibinfo {title} {Quantum Phase Transitions}}},\ \bibinfo {edition} {2nd}\
	ed.\ (\bibinfo  {publisher} {Cambridge University Press},\ \bibinfo {year}
	{2011})\BibitemShut {NoStop}%
	\bibitem [{\citenamefont {Yasuda}\ \emph {et~al.}(2005)\citenamefont {Yasuda},
		\citenamefont {Todo}, \citenamefont {Hukushima}, \citenamefont {Alet},
		\citenamefont {Keller}, \citenamefont {Troyer},\ and\ \citenamefont
		{Takayama}}]{QMC}%
	\BibitemOpen
	\bibfield  {author} {\bibinfo {author} {\bibfnamefont {C.}~\bibnamefont
			{Yasuda}}, \bibinfo {author} {\bibfnamefont {S.}~\bibnamefont {Todo}},
		\bibinfo {author} {\bibfnamefont {K.}~\bibnamefont {Hukushima}}, \bibinfo
		{author} {\bibfnamefont {F.}~\bibnamefont {Alet}}, \bibinfo {author}
		{\bibfnamefont {M.}~\bibnamefont {Keller}}, \bibinfo {author} {\bibfnamefont
			{M.}~\bibnamefont {Troyer}},\ and\ \bibinfo {author} {\bibfnamefont
			{H.}~\bibnamefont {Takayama}},\ }\bibfield  {title} {\bibinfo {title} {N\'eel
			temperature of quasi-low-dimensional {Heisenberg} antiferromagnets},\ }\href
	{https://doi.org/10.1103/PhysRevLett.94.217201} {\bibfield  {journal}
		{\bibinfo  {journal} {Phys. Rev. Lett.}\ }\textbf {\bibinfo {volume} {94}},\
		\bibinfo {pages} {217201} (\bibinfo {year} {2005})}\BibitemShut {NoStop}%
	\bibitem [{\citenamefont {Schulz}(1996)}]{schulz}%
	\BibitemOpen
	\bibfield  {author} {\bibinfo {author} {\bibfnamefont {H.~J.}\ \bibnamefont
			{Schulz}},\ }\bibfield  {title} {\bibinfo {title} {Dynamics of coupled
			quantum spin chains},\ }\href {https://doi.org/10.1103/PhysRevLett.77.2790}
	{\bibfield  {journal} {\bibinfo  {journal} {Phys. Rev. Lett.}\ }\textbf
		{\bibinfo {volume} {77}},\ \bibinfo {pages} {2790} (\bibinfo {year}
		{1996})}\BibitemShut {NoStop}%
	\bibitem [{\citenamefont {Lancaster}\ \emph {et~al.}(2006)\citenamefont
		{Lancaster}, \citenamefont {Blundell}, \citenamefont {Brooks}, \citenamefont
		{Baker}, \citenamefont {Pratt}, \citenamefont {Manson}, \citenamefont
		{Landee},\ and\ \citenamefont {Baines}}]{cupyz}%
	\BibitemOpen
	\bibfield  {author} {\bibinfo {author} {\bibfnamefont {T.}~\bibnamefont
			{Lancaster}}, \bibinfo {author} {\bibfnamefont {S.~J.}\ \bibnamefont
			{Blundell}}, \bibinfo {author} {\bibfnamefont {M.~L.}\ \bibnamefont
			{Brooks}}, \bibinfo {author} {\bibfnamefont {P.~J.}\ \bibnamefont {Baker}},
		\bibinfo {author} {\bibfnamefont {F.~L.}\ \bibnamefont {Pratt}}, \bibinfo
		{author} {\bibfnamefont {J.~L.}\ \bibnamefont {Manson}}, \bibinfo {author}
		{\bibfnamefont {C.~P.}\ \bibnamefont {Landee}},\ and\ \bibinfo {author}
		{\bibfnamefont {C.}~\bibnamefont {Baines}},\ }\bibfield  {title} {\bibinfo
		{title} {Magnetic order in the quasi-one-dimensional spin-$1/2$ molecular
			chain compound copper pyrazine dinitrate},\ }\href
	{https://doi.org/10.1103/PhysRevB.73.020410} {\bibfield  {journal} {\bibinfo
			{journal} {Phys. Rev. B}\ }\textbf {\bibinfo {volume} {73}},\ \bibinfo
		{pages} {020410(R)} (\bibinfo {year} {2006})}\BibitemShut {NoStop}%
	\bibitem [{\citenamefont {Hammar}\ \emph {et~al.}(1999)\citenamefont {Hammar},
		\citenamefont {Stone}, \citenamefont {Reich}, \citenamefont {Broholm},
		\citenamefont {Gibson}, \citenamefont {Turnbull}, \citenamefont {Landee},\
		and\ \citenamefont {Oshikawa}}]{hammar}%
	\BibitemOpen
	\bibfield  {author} {\bibinfo {author} {\bibfnamefont {P.~R.}\ \bibnamefont
			{Hammar}}, \bibinfo {author} {\bibfnamefont {M.~B.}\ \bibnamefont {Stone}},
		\bibinfo {author} {\bibfnamefont {D.~H.}\ \bibnamefont {Reich}}, \bibinfo
		{author} {\bibfnamefont {C.}~\bibnamefont {Broholm}}, \bibinfo {author}
		{\bibfnamefont {P.~J.}\ \bibnamefont {Gibson}}, \bibinfo {author}
		{\bibfnamefont {M.~M.}\ \bibnamefont {Turnbull}}, \bibinfo {author}
		{\bibfnamefont {C.~P.}\ \bibnamefont {Landee}},\ and\ \bibinfo {author}
		{\bibfnamefont {M.}~\bibnamefont {Oshikawa}},\ }\bibfield  {title} {\bibinfo
		{title} {Characterization of a quasi-one-dimensional spin-1/2 magnet which is
			gapless and paramagnetic for
			{$g{\ensuremath{\mu}}_{B}H\ensuremath{\lesssim}J$} and
			{${k}_{B}T\ensuremath{\ll}J$}},\ }\href
	{https://doi.org/10.1103/PhysRevB.59.1008} {\bibfield  {journal} {\bibinfo
			{journal} {Phys. Rev. B}\ }\textbf {\bibinfo {volume} {59}},\ \bibinfo
		{pages} {1008} (\bibinfo {year} {1999})}\BibitemShut {NoStop}%
	\bibitem [{\citenamefont {Liu}\ \emph {et~al.}(2019)\citenamefont {Liu},
		\citenamefont {Kittaka}, \citenamefont {Johnson}, \citenamefont {Lancaster},
		\citenamefont {Singleton}, \citenamefont {Sakakibara}, \citenamefont
		{Kohama}, \citenamefont {van Tol}, \citenamefont {Ardavan}, \citenamefont
		{Williams}, \citenamefont {Blundell}, \citenamefont {Manson}, \citenamefont
		{Manson},\ and\ \citenamefont {Goddard}}]{liu}%
	\BibitemOpen
	\bibfield  {author} {\bibinfo {author} {\bibfnamefont {J.}~\bibnamefont
			{Liu}}, \bibinfo {author} {\bibfnamefont {S.}~\bibnamefont {Kittaka}},
		\bibinfo {author} {\bibfnamefont {R.~D.}\ \bibnamefont {Johnson}}, \bibinfo
		{author} {\bibfnamefont {T.}~\bibnamefont {Lancaster}}, \bibinfo {author}
		{\bibfnamefont {J.}~\bibnamefont {Singleton}}, \bibinfo {author}
		{\bibfnamefont {T.}~\bibnamefont {Sakakibara}}, \bibinfo {author}
		{\bibfnamefont {Y.}~\bibnamefont {Kohama}}, \bibinfo {author} {\bibfnamefont
			{J.}~\bibnamefont {van Tol}}, \bibinfo {author} {\bibfnamefont
			{A.}~\bibnamefont {Ardavan}}, \bibinfo {author} {\bibfnamefont {B.~H.}\
			\bibnamefont {Williams}}, \bibinfo {author} {\bibfnamefont {S.~J.}\
			\bibnamefont {Blundell}}, \bibinfo {author} {\bibfnamefont {Z.~E.}\
			\bibnamefont {Manson}}, \bibinfo {author} {\bibfnamefont {J.~L.}\
			\bibnamefont {Manson}},\ and\ \bibinfo {author} {\bibfnamefont {P.~A.}\
			\bibnamefont {Goddard}},\ }\bibfield  {title} {\bibinfo {title}
		{Unconventional field-induced spin gap in an {$S=1/2$} chiral staggered
			chain},\ }\href {https://doi.org/10.1103/PhysRevLett.122.057207} {\bibfield
		{journal} {\bibinfo  {journal} {Phys. Rev. Lett.}\ }\textbf {\bibinfo
			{volume} {122}},\ \bibinfo {pages} {057207} (\bibinfo {year}
		{2019})}\BibitemShut {NoStop}%
	\bibitem [{\citenamefont {Cordes}\ and\ \citenamefont {Rogers}(2007)}]{cordes}%
	\BibitemOpen
	\bibfield  {author} {\bibinfo {author} {\bibfnamefont {B.}~\bibnamefont
			{Cordes}, \bibfnamefont {David}}\ and\ \bibinfo {author} {\bibfnamefont
			{R.~D.}\ \bibnamefont {Rogers}},\ }\bibfield  {title} {\bibinfo {title}
		{Enantiomorphic helical coordination polymers of
			\{[M(pyrimidine)(OH$_2$)$_4$][SiF$_6$]{\textperiodcentered}H$_2$O\}$_\infty$
			({M} = {Co}$^{2+}$, {Cu}$^{2+}$, {Zn}$^{2+}$)},\ }\href
	{https://doi.org/10.1021/cg070462p} {\bibfield  {journal} {\bibinfo
			{journal} {Cryst. Growth Des}\ }\textbf {\bibinfo {volume} {7}},\ \bibinfo
		{pages} {1943} (\bibinfo {year} {2007})}\BibitemShut {NoStop}%
	\bibitem [{\citenamefont {Jin}\ and\ \citenamefont {Starykh}(2017)}]{starykh}%
	\BibitemOpen
	\bibfield  {author} {\bibinfo {author} {\bibfnamefont {W.}~\bibnamefont
			{Jin}}\ and\ \bibinfo {author} {\bibfnamefont {O.~A.}\ \bibnamefont
			{Starykh}},\ }\bibfield  {title} {\bibinfo {title} {Phase diagram of weakly
			coupled {Heisenberg} spin chains subject to a uniform
			{Dzyaloshinskii}-{Moriya} interaction},\ }\href
	{https://doi.org/10.1103/PhysRevB.95.214404} {\bibfield  {journal} {\bibinfo
			{journal} {Phys. Rev. B}\ }\textbf {\bibinfo {volume} {95}},\ \bibinfo
		{pages} {214404} (\bibinfo {year} {2017})}\BibitemShut {NoStop}%
	\bibitem [{\citenamefont {Xiao}\ \emph {et~al.}(2015)\citenamefont {Xiao},
		\citenamefont {M\"oller}, \citenamefont {Lancaster}, \citenamefont
		{Williams}, \citenamefont {Pratt}, \citenamefont {Blundell}, \citenamefont
		{Ceresoli}, \citenamefont {Barton},\ and\ \citenamefont {Manson}}]{xiao}%
	\BibitemOpen
	\bibfield  {author} {\bibinfo {author} {\bibfnamefont {F.}~\bibnamefont
			{Xiao}}, \bibinfo {author} {\bibfnamefont {J.~S.}\ \bibnamefont {M\"oller}},
		\bibinfo {author} {\bibfnamefont {T.}~\bibnamefont {Lancaster}}, \bibinfo
		{author} {\bibfnamefont {R.~C.}\ \bibnamefont {Williams}}, \bibinfo {author}
		{\bibfnamefont {F.~L.}\ \bibnamefont {Pratt}}, \bibinfo {author}
		{\bibfnamefont {S.~J.}\ \bibnamefont {Blundell}}, \bibinfo {author}
		{\bibfnamefont {D.}~\bibnamefont {Ceresoli}}, \bibinfo {author}
		{\bibfnamefont {A.~M.}\ \bibnamefont {Barton}},\ and\ \bibinfo {author}
		{\bibfnamefont {J.~L.}\ \bibnamefont {Manson}},\ }\bibfield  {title}
	{\bibinfo {title} {Spin diffusion in the low-dimensional molecular quantum
			{Heisenberg} antiferromagnet
			$\mathrm{Cu}(\mathrm{pyz}){({\mathrm{NO}}_{3})}_{2}$ detected with implanted
			muons},\ }\href {https://doi.org/10.1103/PhysRevB.91.144417} {\bibfield
		{journal} {\bibinfo  {journal} {Phys. Rev. B}\ }\textbf {\bibinfo {volume}
			{91}},\ \bibinfo {pages} {144417} (\bibinfo {year} {2015})}\BibitemShut
	{NoStop}%
	\bibitem [{\citenamefont {Lancaster}\ \emph {et~al.}(2018)\citenamefont
		{Lancaster}, \citenamefont {Xiao}, \citenamefont {Huddart}, \citenamefont
		{Williams}, \citenamefont {Pratt}, \citenamefont {Blundell}, \citenamefont
		{Clark}, \citenamefont {Scheuermann}, \citenamefont {Goko}, \citenamefont
		{Ward}, \citenamefont {Manson}, \citenamefont {Rüegg},\ and\ \citenamefont
		{Krämer}}]{spinladder}%
	\BibitemOpen
	\bibfield  {author} {\bibinfo {author} {\bibfnamefont {T.}~\bibnamefont
			{Lancaster}}, \bibinfo {author} {\bibfnamefont {F.}~\bibnamefont {Xiao}},
		\bibinfo {author} {\bibfnamefont {B.~M.}\ \bibnamefont {Huddart}}, \bibinfo
		{author} {\bibfnamefont {R.~C.}\ \bibnamefont {Williams}}, \bibinfo {author}
		{\bibfnamefont {F.~L.}\ \bibnamefont {Pratt}}, \bibinfo {author}
		{\bibfnamefont {S.~J.}\ \bibnamefont {Blundell}}, \bibinfo {author}
		{\bibfnamefont {S.~J.}\ \bibnamefont {Clark}}, \bibinfo {author}
		{\bibfnamefont {R.}~\bibnamefont {Scheuermann}}, \bibinfo {author}
		{\bibfnamefont {T.}~\bibnamefont {Goko}}, \bibinfo {author} {\bibfnamefont
			{S.}~\bibnamefont {Ward}}, \bibinfo {author} {\bibfnamefont {J.~L.}\
			\bibnamefont {Manson}}, \bibinfo {author} {\bibfnamefont {C.}~\bibnamefont
			{Rüegg}},\ and\ \bibinfo {author} {\bibfnamefont {K.~W.}\ \bibnamefont
			{Krämer}},\ }\bibfield  {title} {\bibinfo {title} {Quantum magnetism in
			molecular spin ladders probed with muon-spin spectroscopy},\ }\href
	{https://doi.org/10.1088/1367-2630/aae21a} {\bibfield  {journal} {\bibinfo
			{journal} {New J. Phys.}\ }\textbf {\bibinfo {volume} {20}},\ \bibinfo
		{pages} {103002} (\bibinfo {year} {2018})}\BibitemShut {NoStop}%
	\bibitem [{\citenamefont {Devreux}\ \emph {et~al.}(1974)\citenamefont
		{Devreux}, \citenamefont {Boucher},\ and\ \citenamefont
		{Nechtschein}}]{devreux}%
	\BibitemOpen
	\bibfield  {author} {\bibinfo {author} {\bibfnamefont {F.}~\bibnamefont
			{Devreux}}, \bibinfo {author} {\bibfnamefont {J.-P.}\ \bibnamefont
			{Boucher}},\ and\ \bibinfo {author} {\bibfnamefont {M.}~\bibnamefont
			{Nechtschein}},\ }\bibfield  {title} {\bibinfo {title} {Temps de relaxation
			nucl\'eaire {T1D} et {T1$\rho$} en pr\'esence de mouvement de spins
			\'electroniques},\ }\href {https://doi.org/10.1051/jphys:01974003503027100}
	{\bibfield  {journal} {\bibinfo  {journal} {J. Phys. France}\ }\textbf
		{\bibinfo {volume} {35}},\ \bibinfo {pages} {271} (\bibinfo {year}
		{1974})}\BibitemShut {NoStop}%
	\bibitem [{\citenamefont {Pratt}\ \emph {et~al.}(2006)\citenamefont {Pratt},
		\citenamefont {Blundell}, \citenamefont {Lancaster}, \citenamefont {Baines},\
		and\ \citenamefont {Takagi}}]{pratt}%
	\BibitemOpen
	\bibfield  {author} {\bibinfo {author} {\bibfnamefont {F.~L.}\ \bibnamefont
			{Pratt}}, \bibinfo {author} {\bibfnamefont {S.~J.}\ \bibnamefont {Blundell}},
		\bibinfo {author} {\bibfnamefont {T.}~\bibnamefont {Lancaster}}, \bibinfo
		{author} {\bibfnamefont {C.}~\bibnamefont {Baines}},\ and\ \bibinfo {author}
		{\bibfnamefont {S.}~\bibnamefont {Takagi}},\ }\bibfield  {title} {\bibinfo
		{title} {Low-temperature spin diffusion in a highly ideal {$S=\frac{1}{2}$}
			{Heisenberg} antiferromagnetic chain studied by muon spin relaxation},\
	}\href {https://doi.org/10.1103/PhysRevLett.96.247203} {\bibfield  {journal}
		{\bibinfo  {journal} {Phys. Rev. Lett.}\ }\textbf {\bibinfo {volume} {96}},\
		\bibinfo {pages} {247203} (\bibinfo {year} {2006})}\BibitemShut {NoStop}%
	\bibitem [{\citenamefont {Mizoguchi}(1995)}]{mizoguchi}%
	\BibitemOpen
	\bibfield  {author} {\bibinfo {author} {\bibfnamefont {K.}~\bibnamefont
			{Mizoguchi}},\ }\bibfield  {title} {\bibinfo {title} {Spin dynamics study in
			conducting polymers by magnetic resonance},\ }\href
	{https://doi.org/10.7567/jjap.34.1} {\bibfield  {journal} {\bibinfo
			{journal} {Jpn. J. Appl. Phys.}\ }\textbf {\bibinfo {volume} {34}},\ \bibinfo
		{pages} {1} (\bibinfo {year} {1995})}\BibitemShut {NoStop}%
	\bibitem [{\citenamefont {Lancaster}\ \emph {et~al.}(2012)\citenamefont
		{Lancaster}, \citenamefont {Baker}, \citenamefont {Pratt}, \citenamefont
		{Blundell}, \citenamefont {Hayes},\ and\ \citenamefont
		{Prabhakaran}}]{e184404}%
	\BibitemOpen
	\bibfield  {author} {\bibinfo {author} {\bibfnamefont {T.}~\bibnamefont
			{Lancaster}}, \bibinfo {author} {\bibfnamefont {P.~J.}\ \bibnamefont
			{Baker}}, \bibinfo {author} {\bibfnamefont {F.~L.}\ \bibnamefont {Pratt}},
		\bibinfo {author} {\bibfnamefont {S.~J.}\ \bibnamefont {Blundell}}, \bibinfo
		{author} {\bibfnamefont {W.}~\bibnamefont {Hayes}},\ and\ \bibinfo {author}
		{\bibfnamefont {D.}~\bibnamefont {Prabhakaran}},\ }\bibfield  {title}
	{\bibinfo {title} {Persistent dynamics in the {$S=1/2$} quasi-one-dimensional
			chain compound {Rb}$_{4}${Cu}({Mo}{O}$_{4}$)$_{3}$ probed with muon-spin
			relaxation},\ }\href {https://doi.org/10.1103/PhysRevB.85.184404} {\bibfield
		{journal} {\bibinfo  {journal} {Phys. Rev. B}\ }\textbf {\bibinfo {volume}
			{85}},\ \bibinfo {pages} {184404} (\bibinfo {year} {2012})}\BibitemShut
	{NoStop}%
	\bibitem [{\citenamefont {Maeter}\ \emph {et~al.}(2013)\citenamefont {Maeter},
		\citenamefont {Zvyagin}, \citenamefont {Luetkens}, \citenamefont {Pascua},
		\citenamefont {Shermadini}, \citenamefont {Saint-Martin}, \citenamefont
		{Revcolevschi}, \citenamefont {Hess}, \citenamefont {Büchner},\ and\
		\citenamefont {Klauss}}]{maeter}%
	\BibitemOpen
	\bibfield  {author} {\bibinfo {author} {\bibfnamefont {H.}~\bibnamefont
			{Maeter}}, \bibinfo {author} {\bibfnamefont {A.~A.}\ \bibnamefont {Zvyagin}},
		\bibinfo {author} {\bibfnamefont {H.}~\bibnamefont {Luetkens}}, \bibinfo
		{author} {\bibfnamefont {G.}~\bibnamefont {Pascua}}, \bibinfo {author}
		{\bibfnamefont {Z.}~\bibnamefont {Shermadini}}, \bibinfo {author}
		{\bibfnamefont {R.}~\bibnamefont {Saint-Martin}}, \bibinfo {author}
		{\bibfnamefont {A.}~\bibnamefont {Revcolevschi}}, \bibinfo {author}
		{\bibfnamefont {C.}~\bibnamefont {Hess}}, \bibinfo {author} {\bibfnamefont
			{B.}~\bibnamefont {Büchner}},\ and\ \bibinfo {author} {\bibfnamefont
			{H.-H.}\ \bibnamefont {Klauss}},\ }\bibfield  {title} {\bibinfo {title} {Low
			temperature ballistic spin transport in the {$S$}= 1/2 antiferromagnetic
			{Heisenberg} chain compound {SrCuO}$_2$},\ }\href
	{https://doi.org/10.1088/0953-8984/25/36/365601} {\bibfield  {journal}
		{\bibinfo  {journal} {J. Phys. Condens. Matter}\ }\textbf {\bibinfo {volume}
			{25}},\ \bibinfo {pages} {365601} (\bibinfo {year} {2013})}\BibitemShut
	{NoStop}%
	\bibitem [{\citenamefont {Thurber}\ \emph {et~al.}(2001)\citenamefont
		{Thurber}, \citenamefont {Hunt}, \citenamefont {Imai},\ and\ \citenamefont
		{Chou}}]{thurber}%
	\BibitemOpen
	\bibfield  {author} {\bibinfo {author} {\bibfnamefont {K.~R.}\ \bibnamefont
			{Thurber}}, \bibinfo {author} {\bibfnamefont {A.~W.}\ \bibnamefont {Hunt}},
		\bibinfo {author} {\bibfnamefont {T.}~\bibnamefont {Imai}},\ and\ \bibinfo
		{author} {\bibfnamefont {F.~C.}\ \bibnamefont {Chou}},\ }\bibfield  {title}
	{\bibinfo {title} {$^{17}${O} {NMR} study of $\mathit{q}=0$ spin excitations
			in a nearly ideal $\mathit{S}=\frac{1}{2}$ {1D} {Heisenberg} antiferromagnet,
			{Sr}$_2${Cu}{O}$_3$, up to 800 {K}},\ }\href
	{https://doi.org/10.1103/PhysRevLett.87.247202} {\bibfield  {journal}
		{\bibinfo  {journal} {Phys. Rev. Lett.}\ }\textbf {\bibinfo {volume} {87}},\
		\bibinfo {pages} {247202} (\bibinfo {year} {2001})}\BibitemShut {NoStop}%
	\bibitem [{\citenamefont {Sirker}\ \emph {et~al.}(2009)\citenamefont {Sirker},
		\citenamefont {Pereira},\ and\ \citenamefont {Affleck}}]{sirker}%
	\BibitemOpen
	\bibfield  {author} {\bibinfo {author} {\bibfnamefont {J.}~\bibnamefont
			{Sirker}}, \bibinfo {author} {\bibfnamefont {R.~G.}\ \bibnamefont
			{Pereira}},\ and\ \bibinfo {author} {\bibfnamefont {I.}~\bibnamefont
			{Affleck}},\ }\bibfield  {title} {\bibinfo {title} {Diffusion and ballistic
			transport in one-dimensional quantum systems},\ }\href
	{https://doi.org/10.1103/PhysRevLett.103.216602} {\bibfield  {journal}
		{\bibinfo  {journal} {Phys. Rev. Lett.}\ }\textbf {\bibinfo {volume} {103}},\
		\bibinfo {pages} {216602} (\bibinfo {year} {2009})}\BibitemShut {NoStop}%
	\bibitem [{\citenamefont {Tiegel}\ \emph {et~al.}(2016)\citenamefont {Tiegel},
		\citenamefont {Honecker}, \citenamefont {Pruschke}, \citenamefont
		{Ponomaryov}, \citenamefont {Zvyagin}, \citenamefont {Feyerherm},\ and\
		\citenamefont {Manmana}}]{tiegel}%
	\BibitemOpen
	\bibfield  {author} {\bibinfo {author} {\bibfnamefont {A.~C.}\ \bibnamefont
			{Tiegel}}, \bibinfo {author} {\bibfnamefont {A.}~\bibnamefont {Honecker}},
		\bibinfo {author} {\bibfnamefont {T.}~\bibnamefont {Pruschke}}, \bibinfo
		{author} {\bibfnamefont {A.}~\bibnamefont {Ponomaryov}}, \bibinfo {author}
		{\bibfnamefont {S.~A.}\ \bibnamefont {Zvyagin}}, \bibinfo {author}
		{\bibfnamefont {R.}~\bibnamefont {Feyerherm}},\ and\ \bibinfo {author}
		{\bibfnamefont {S.~R.}\ \bibnamefont {Manmana}},\ }\bibfield  {title}
	{\bibinfo {title} {Dynamical properties of the sine-{Gordon} quantum spin
			magnet {Cu}-{PM} at zero and finite temperature},\ }\href
	{https://doi.org/10.1103/PhysRevB.93.104411} {\bibfield  {journal} {\bibinfo
			{journal} {Phys. Rev. B}\ }\textbf {\bibinfo {volume} {93}},\ \bibinfo
		{pages} {104411} (\bibinfo {year} {2016})}\BibitemShut {NoStop}%
\end{thebibliography}
\end{document}